\documentclass{iaus}
\usepackage{graphicx}

\title[Convective Overshooting] 
{Effects of Rotation and Input Energy Flux on Convective Overshooting}

\author[K\"apyl\"a, Korpi, Stix \& Tuominen]   
{Petri J. K\"apyl\"a$^{1,2}$, M.J. Korpi$^1$, M. Stix$^2$ \break \and I. Tuominen$^1$}

\affiliation{$^1$Observatory, P.O. Box 14, FI-00014 University of 
Helsinki, Finland \\[\affilskip]
$^2$Kiepenheuer-Institut f\"ur Sonnenphysik, Sch\"oneckstrasse 6, D-79104 Freiburg, Germany \break email: pkapyla@kis.uni-freiburg.de}

\pubyear{2007}
\volume{239}  
\pagerange{119--126}
\date{?? and in revised form ??}
\jname{Convection in Astrophysics}
\editors{F. Kupka, I. W. Roxburgh \& K.L. Chan eds.}
\begin{document}

\maketitle

\begin{abstract}
We study convective overshooting by means of local 3D convection
calculations. Using a mixing length model of the solar convection zone
(CZ) as a guide, we determine the Coriolis number (Co), which is the
inverse of the Rossby number, to be of the order of ten or larger at
the base of the solar CZ. Therefore we perform convection calculations
in the range Co = 0...10 and interpret the value of Co realised in the
calculation to represent a depth in the solar CZ. In order to study
the dependence on rotation, we compute the mixing length parameters
$\alpha_{\rm T}$ and $\alpha_{\rm u}$ relating the temperature and
velocity fluctuations, respectively, to the mean thermal
stratification. We find that the mixing length parameters for the
rapid rotation case, corresponding to the base of the solar CZ, are
3-5 times smaller than in the nonrotating case. Introducing such
depth-dependent $\alpha$ into a solar structure model employing a
non-local mixing length formalism results in overshooting which is
approximately proportional to $\alpha$ at the base of the CZ. Although
overshooting is reduced due to the reduced $\alpha$, a discrepancy
with helioseismology remains due to the steep transition to the
radiative temperature gradient.

In comparison to the mixing length models the transition at the base
of the CZ is much gentler in the 3D models. It was suggested recently
(Rempel \cite{Rempel2004}) that this discrepancy is due to the
significantly larger (up to seven orders of magnitude) input energy
flux in the 3D models in comparison to the Sun and solar models, and
that the 3D calculations should be able to approach the mixing length
regime if the input energy flux is decreased by a moderate amount. We
present results from local convection calculations which support this
conjecture.  
\end{abstract}

\firstsection 
\section{Introduction}
Convection poses a difficult problem for stellar structure
modelling. One-dimensional stellar structure models require a
parameterization of convection in order to be able to yield the
thermal stratification within the CZ. Since convection in the stellar
envelopes is in general highly efficient, the stratification is close
to adiabatic in much of the CZ and thus a detailed description of
convection is not needed there. The most often used way to
parameterize convection is to use the mixing length concept in a form
introduced by \cite{Vitense1953}; see also \cite{BoehmVitense1958},
which considers convective elements to lose their identity after
rising (or descending) the so-called mixing length which is
proportional to the local pressure scale height, i.e.
\begin{equation}
l = \alpha H_{\rm p}\;.\label{equ:mlt}
\end{equation}
Using this basic assumption, it is possible to derive equations that
relate the velocity and temperature fluctuations to the mean
stratification, and thus, to compute the convective energy flux (see,
e.g. Chapter 6 of Stix \cite{Stix2002}).

\begin{figure}
\begin{centering}
\includegraphics[height=3.5in,width=3.5in]{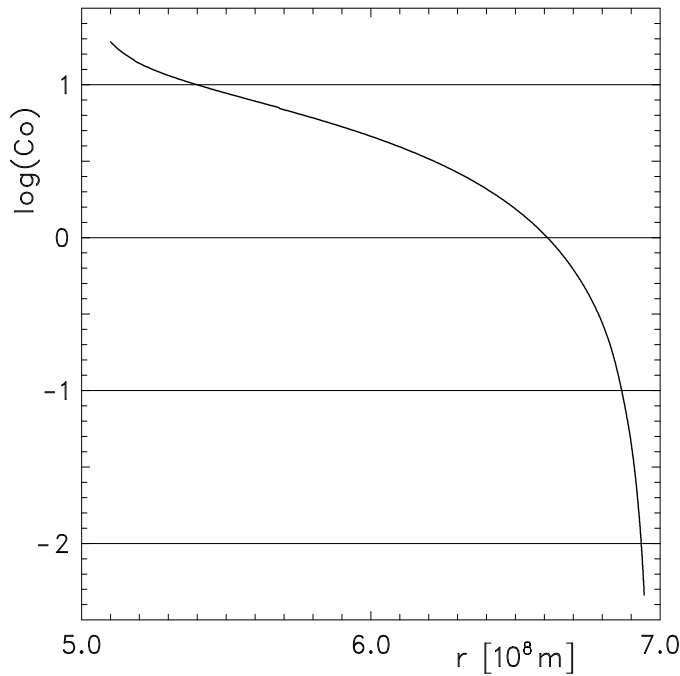}
  \caption{Coriolis number in the solar convection zone according to
    (\ref{equ:Co}).}\label{fig:comlt}
\end{centering}
\end{figure}

Among other conceptual problems, the mixing length formalism neglects
the effects of rotation although convection can be significantly
influenced by it in regions of the CZ where the turnover time is
longer than the rotation period. In a recent study, K\"apyl\"a et
al. (\cite{Kaepylaeea2005}) estimated the Coriolis number, which is
the inverse of the Rossby number, in the solar CZ from
\begin{equation}
{\rm Co} = {\rm Ro}^{-1} = 2\,\Omega_\odot \tau = 2\,\Omega_\odot \alpha H_{\rm p}/u\;, \label{equ:Co}
\end{equation}
where $\alpha = 1.66$, $\Omega_\odot = 2.6 \cdot 10^{-6}$\,s$^{-1}$,
and $u$ the convective velocity obtained using a local mixing length
model. This relation gives values of the order of $10^{-3}$ near the
solar surface and of the order of ten or larger near the base of the
CZ (see figure \ref{fig:comlt}). We compute the mixing length
parameters relating the velocity and temperature fluctuations to the
mean stratification from local 3D convection calculations in the range
${\rm Co} = 0 \ldots 10$, which coincides with the range expected in
the Sun. Thus it is possible to probe the influence of rotation on the
mixing length relations and take the effect implicitly into account in
a solar model employing a non-local formulation of the mixing length
concept in order to study overshooting below the CZ.

There is a striking difference between the almost adiabatic
overshooting with a very sharp transition to the radiative gradient
seen in non-local mixing length models in comparison to the much more
subadiabatic and smoother overshooting seen in 3D convection
models. In a recent paper Rempel (\cite{Rempel2004}) suggested that
this discrepancy arises due to the fact that the two models are simply
working in different parameter regimes in the sense that the input
energy flux in the 3D models is usually up to $10^7$ times larger in
comparison to the non-local mixing length models and the Sun. We have
performed 3D numerical calculations in which we decrease the input
flux by two orders of magnitude in order to study this effect in more
detail.

The remainder of the paper is organised as follows:
in~\S\,\ref{sec:nummodel} a brief description of the model is given
and in~\S\,\ref{subsec:mltrot} and \S\,\ref{subsec:inpute} the results
concerning the effects of rotation and input energy flux on convective
overshooting are presented. Finally, the main results of the study and
remaining problems are summarised in~\S\,\ref{sec:concl}.

\section{Numerical model}\label{sec:nummodel}
We use the same model as that described in K\"apyl\"a et
al. (\cite{Kaepylaeea2005}, \cite{Kaepylaeea2006}). The computational
domain is a rectangular box situated at a latitude $\Theta$, in which
case the rotation vector is represented by $\boldsymbol{\Omega} =
\Omega_0 (\cos \Theta, 0, -\sin \Theta)$. In the present study the
calculations with rotation are performed at the south pole,
i.e. $\Theta = -90^{\rm o}$. The fluid obeys the ideal gas law and
radiation is taken into account only via the diffusion approximation.

In contrast to many earlier studies (e.g. Brummell et
al. \cite{Brummellea2002}) we do not use a piecewise polytropic
stratification which implies that the thermal conductivity behaves
like a step function. Instead, we use a smoothly varying
stratification where the logarithmic temperature gradient is computed
from
\begin{equation}
     \nabla = \nabla_{\rm 3} + \frac{1}{2} \{\tanh [4(z_{\rm m} - z)] +
     1\}\Delta \nabla\;,\label{equ:iniconf}
\end{equation}
where $\nabla_3 = 0.15$ is the gradient at the bottom, $\Delta \nabla
= \nabla_{\rm CZ} - \nabla_3$ the difference between the gradient in
the unstable layer and the applied gradient, and $z_{\rm m}$ the
inflection point of the tanh-function, calculated so that $\nabla =
\nabla_{\rm ad}$ in the initial state at the base of the convectively
unstable region at $z/d = 1$.

In order to regulate the input energy flux we split the heat
conduction term, $\partial_t e = \ldots + \Gamma_{\rm cond}$, in the
internal energy equation into two parts
\begin{eqnarray}
  \Gamma_{\rm cond} = \nabla \cdot [\kappa_{\rm t} \nabla (e - \overline{e}) + \kappa_{\rm h} \nabla \overline{e}]\;,
\end{eqnarray}
where the first term acts only on the fluctuations and the latter only
on the mean, i.e. horizontally averaged stratification, and where $e =
c_{\rm V} T$. Thus $\kappa_{\rm t}$ and $\kappa_{\rm h}$ can be
considered as the turbulent and radiative conductivities, which
satisfy $\kappa_{\rm t} \gg \kappa_{\rm h}$ in real stars. We define
the conductivies as
\begin{eqnarray}
  \kappa_{\rm t} &=& \gamma \rho \chi_0\;, \\
  \kappa_{\rm h} &=& \frac{(\gamma-1) F_{\rm b}}{g \nabla} \;, \label{equ:kappah}
\end{eqnarray}
where $\chi_0$ is the reference value of the thermal diffusivity,
computed from ${\rm Pr} = \nu/\chi_0$, where ${\rm Pr} = 0.4$ is the
Prandtl number and $\nu$ the kinematic viscosity. $F_{\rm b}$ is the
input energy flux, $g$ the constant gravitational acceleration, and
$\nabla$ the mean logarithmic temperature gradient given by
(\ref{equ:iniconf}) (for more details, see K\"apyl\"a et
al. \cite{Kaepylaeea2006}).

\section{Results}\label{sec:results}

\subsection{Effects of rotation on mixing length relations}\label{subsec:mltrot}
\begin{figure}
\begin{centering}
\includegraphics[height=2.25in,width=5in]{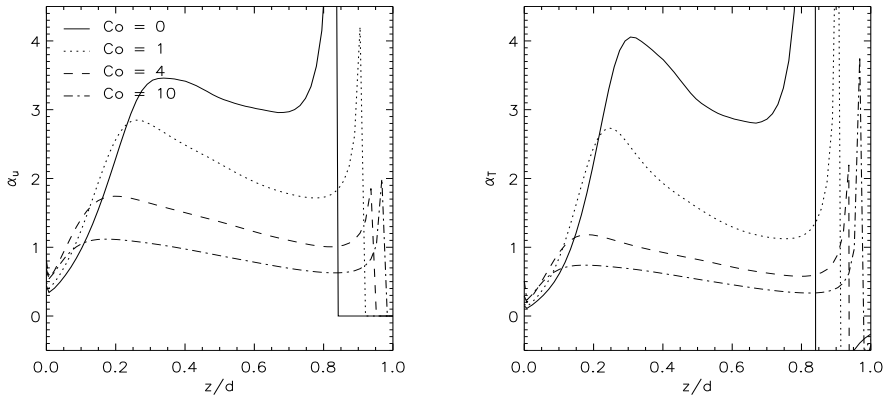}
  \caption{Mixing length parameters according to (\ref{fig:alphau})
    and (\ref{fig:alphat}).}\label{fig:alphas}
\end{centering}
\end{figure}

Using the basic assumption of the mixing length concept,
(\ref{equ:mlt}), it is possible to derive equations that relate the
velocity and temperature fluctuations to the mean thermal
stratification
\begin{eqnarray}
\overline{u'^2_z} = \frac{\alpha_{\rm u}^2 H_{\rm p} g}{8} (\nabla - \nabla_{\rm ad})\;, \label{fig:alphau} \\
\overline{T'^2} = \frac{\alpha_{\rm T}}{2}(\nabla - \nabla_{\rm ad}) \overline{T}\;,\label{fig:alphat}
\end{eqnarray}
where the bars denote horizontal averaging, primes the fluctuation,
and $\nabla_{\rm ad} = (\gamma - 1)/\gamma = 0.4$, where $\gamma =
c_{\rm P}/c_{\rm V} = 5/3$. Furthermore, adiabatic variation within
the convective elements is assumed. Figure \ref{fig:alphas} shows the
results for calculations made at the south pole with approximate
Coriolis numbers 0 (no rotation), 1, 4, and 10. It is clear that
increasing rotation reduces the convective efficiency and thus the
mixing length parameters. If the rotation dependence is intepreted as
a depth dependence in the solar CZ, the reduction of the mixing length
$\alpha$ can be taken into account in solar models as an implicit way
of incorporating the effects of rotation on convection. The main
effect of reduced $\alpha$ near the base of the CZ is the reduction of
the overshooting when a non-local version of the mixing length model
is used (K\"apyl\"a et al. \cite{Kaepylaeea2005}). The overshooting
depth is approximately proportional to the mixing length at the base
of the CZ.

\subsection{Effects of input energy flux}\label{subsec:inpute}
The non-local mixing length models tend to produce a quasi-adiabatic
overshoot region with a very sharp transition to the radiative
gradient below the CZ (see e.g. figure 9 of K\"apyl\"a et
al. \cite{Kaepylaeea2005}). Although rotational effects may be able to
alleviate the situation, the sharp transition should still show up in
the helioseismic inversions of the solar internal structure. On the
other hand, numerical 3D calculations always tend to produce
overshooting with a much gentler transition to the radiative gradient
(e.g. Brummell et al. \cite{Brummellea2002}). The main difference
between these two approaches is that whereas in the mixing lenght
models the input energy flux is the solar flux, i.e. $f = F_\odot/\rho
c_{\rm s}^3 \approx 10^{-11}$ in the deep layers of the CZ, the 3D
models need a much higher flux (up to $10^7$ times) in order to bring
the thermal relaxation time closer to the dynamical time scale.

\begin{figure}
\begin{centering}
\includegraphics[height=3.5in,width=5in]{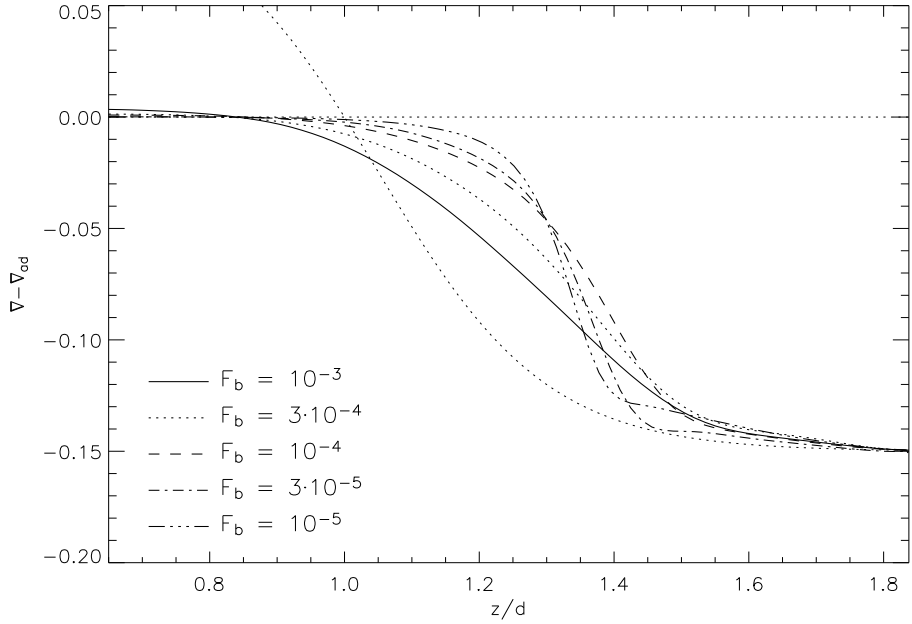}
  \caption{Superadiabatic temperature gradient $\delta = \nabla -
    \nabla_{\rm ad}$ as a function of the input energy flux $F_{\rm
      b}$. The input flux is given in units of $\rho_0 (gd)^{3/2}$,
    see K\"apyl\"a et al. (\cite{Kaepylaeea2006}) for the details. In
    the present models $\rho c_s^3 \approx \rho_0 (gd)^{3/2}$ at the
    base of the convectively unstable region leading to dimensionless
    flux of $f = F_{\rm b}/\rho c_s^3 \approx \mathcal{O}(F_{\rm
      b})$. The thin dotted curve shows the temperature gradient if
    the total flux would be transported by radiative
    diffusion.}\label{fig:superaflux}
\end{centering}
\end{figure}

Recently, Rempel (\cite{Rempel2004}) suggested that if the input
energy flux in 3D calculations was reduced by a moderate amount, the
mixing length regime could be approached. Our results (see figure
\ref{fig:superaflux}) support this conjecture. When the input energy
flux is reduced by a factor of $10^2$ it is seen that the overshooting
depth decreases as the average velocities are reduced, and that the
transition to the radiative gradient becomes significantly steeper. If
these results are taken at face value it would seem difficult to avoid
the quasi-adiabatic overshoot region with steep transition at the base
of the solar CZ if extrapolated to the solar regime which is still
five orders of magnitude away in terms of the input energy flux. One
must, however, bear in mind that in the present models the
convectively unstable region spans only little over two pressure scale
heights so the effects of compressibility are likely to be weak in
comparison to the Sun, affecting the filling factor of downflows. The
filling plays a crucial role in the overshoot model of Rempel
(\cite{Rempel2004}), with low values ($\approx 10^{-5}$) being able to
produce overshooting with smooth transition also for the solar energy
flux.

\section{Conclusions}\label{sec:concl}
Three dimensional local convection calculations were used to probe the
effects of rotation on mixing length coefficients relating the
temperature and velocity fluctuations to the mean thermal
stratification. It was found that when the rotational influence on the
flow is comparable to that expected in the deep layers of the solar
CZ, the mixing length parameters are reduced by a factor of three to
five. If a depth-dependent mixing length $\alpha$ is introduced into a
solar model, the overshooting at the base of the CZ is reduced
approximately in proportion to the reduction of $\alpha$. Although the
depth of the solar CZ can be correctly reproduced in this way, the
steep transition to the radiative gradient should still be visible in
the helioseismic inversions.

The overshooting in 3D convection calculations is much more
subadiabatic with a smooth transition, which is due to the much higher
input energy flux used to meet the time step constraints. In the
present study we show that decreasing the input flux in 3D
calculations leads to more adiabatic overshooting and sharper
transition at the base of the CZ. Although this result seems to
suggest that the 3D calculations will approach the mixing length
regime when the flux is reduced enough, one must bear in mind that in
the present models the stratification is rather weak in comparison to
the Sun. Thus the effects of compressibility are likely to be
underestimated, and lead to too large filling factor for the
downflows. Furthermore, the spatial size of the downflow plumes is
restricted by the grid size, and it is probable that the filling
factor of downflows further decreases when the resolution is
increased.

To summarize, we stress the point that rotation should be taken into
account in models of convective overshooting since it can exert
considerable influence on convection already in slowly rotating stars
such as the Sun. Furthermore, high resolution numerical studies of
deep convection are needed in order to study whether the convective
overshooting is due to very few strong downflows, producing nearly
adiabatic overshoot region with a steep transition, or whether
downflows of different strengths penetrate into the stable region in a
larger area producing smooth overshooting required by helioseismology.

\begin{acknowledgments}
PJK acknowledges the Finnish graduate school for astronomy and space
physics for financial support. PJK and MJK acknowledge travel support
from the Academy of Finland grant no. 1112020.
\end{acknowledgments}

\begin{discussion}

\discuss{J. Christensen-Dalsgaard}{Comment: Helioseismology shows that
  the sound speed gradient is likely smoother in the Sun than in
  models even without penetration, possibly requiring a subadiabatic
  gradient in the lower parts of the convection zone.}

\discuss{R.F. Stein}{Comment: when you model the base of the
  convection zone, it is very slightly subadiabatic inside the
  convection zone.}

\discuss{P.J. K\"apyl\"a}{This is indeed the case also in our 3D
  models.}

\end{discussion}

\end{document}